\documentclass[letterpaper,  
              ]{jacow}
\makeatletter%
	\ifboolexpr{bool{xetex}}
	 {\renewcommand{\Gin@extensions}{.pdf,%
	                    .png,.jpg,.bmp,.pict,.tif,.psd,.mac,.sga,.tga,.gif,%
	                    .eps,.ps,%
	                    }}{}
\makeatother

\ifboolexpr{bool{xetex} or bool{luatex}} 
 {}                                      
 {\usepackage[utf8]{inputenc}}           

\usepackage[USenglish]{babel}			 

\usepackage[final]{pdfpages}
\usepackage{multirow}
\usepackage{ragged2e}

\ifboolexpr{bool{jacowbiblatex}}%
 {%
  \addbibresource{jacow-test.bib}
  \addbibresource{biblatex-examples.bib}
 }{}
\begin{document}

\title{The Impedance of Flat Metallic Plates with Small Corrugations\thanks{Work supported by the U.S.
Department of Energy, Office of Science, Office of Basic
Energy Sciences, under Contract No. DE-AC02-76SF00515.}}

\setcounter{footnote}{3}
\author{K. Bane\thanks{kbane@slac.stanford.edu}, SLAC National Accelerator Laboratory, Menlo Park, CA, USA }
	
\maketitle

%
   

\section{Introduction}
 The RadiaBeam/SLAC dechirper has been installed and commissioned at the Linac Coherent Light Source (LCLS)~\cite{LCLS}. It consists of two pairs of flat plates with small corrugations, with the beam passing in between. The purpose of a dechirper is to compensate residual energy chirp---energy to longitudinal position correlation---just upstream of the undulator in a linac-based, free electron laser (FEL).
The flat geometry allows for adjustment of strength of the dechirper; having both a horizontal and vertical module allows for cancellation of unavoidable quad wake effects.
The LCLS, however, doesn't generally need extra chirp control; the installed dechirper has, instead, been used more as a fast kicker, to facilitate a two-color mode of operation~\cite{two_color}. 

Analytical formulas for the longitudinal and transverse wakes of a dechirper have been developed to make it easier to do parameter studies and to plan the effective use of the device~\cite{Bane}.
 These formulas are more accurate than results of perturbation calculations of the past~\cite{perturbation}. Note that we are here interested in very short-range wakes/high frequency impedances: the typical rms bunch length $\sigma_z\sim10$~$\mu$m, which implies that the typical frequency of interest is $f\sim c/(2\pi\sigma_z)=5$~THz ($c$ is the speed of light).

In this report we begin by introducing the concepts of wakefield and impedance, particularly in periodic structures and in flat geometry. Then we discuss the corrugated structure as a dechirper, the surface impedance approach to wake calculation, and how explicit analytical formulas for the wakes of the dechirper are obtained. This is followed by comparisons with numerical simulations and with measurements at the LCLS, and finally by a summary.

This report summarizes much of the theoretical work on the dechirper done together with G.~Stupakov, I.~Zagorodnov, and E.~Gjonaj. The comparison with measurements are taken from Ref.~\cite{MarcG}, \cite{Zemella}.
The formulas of this report are given in Gaussian units; to change a wake or impedance to MKS units, one merely multiplies by $Z_0c/(4\pi)$, with $Z_0=377$~$\Omega$.

\section{Wakefields and Impedances}

Let us here limit consideration to periodic structures with boundaries made of metal or dielectrics; {\it e.g.} resistive pipes, dielectric tubes, periodic cavities.
 A driving charge $Q$ passes at the speed of light $c$ through such a structure. A test charge moves on a parallel trajectory, also at speed $c$, but at distance $s$ behind. The longitudinal (point charge) wake, $w(s)$, is the (longitudinal) voltage loss of the test particle per unit charge $Q$ per unit length of structure. 
 Thus, the point charge wake, for structure period $p$, is
\begin{equation}
w(s)=-\frac{1}{Qp}\int_0^p E_z(z,t)\Big|_{t=(s+z)/c}\,dz\ ,
\end{equation}
with $E_z$ the longitudinal electric field, $z$ the longitudinal position, and $t$ the time of the test particle.


 The {\it bunch wake} is given by the convolution of the point charge wake and the (longitudinal) bunch distribution $\lambda(s)$:
 \begin{equation}
 W_\lambda(s)=-\int_0^\infty w(s')\lambda(s-s')\,ds'\ ;\label{bunch_wake_eq}
\end{equation}
note that a value of $W_\lambda<0$ means energy loss by the beam.
 The impedance, for wavenumber $k=2\pi f/c$, is given by the Fourier transform of the wake, $\tilde w(k)$:
\begin{equation}
Z(k)=\tilde w(k)=\frac{1}{c}\int_0^\infty w(s)e^{iks}\,ds\ .
\end{equation}
The transverse wakes and impedances, $w_y(s)$, $Z_y(k)$, which are concerned with the transverse force on the test particle, are defined analogously to the longitudinal ones.


\subsection{Case of Flat Geometry}

 Typically we are interested in wakes of {\it pencil beams}, {\it i.e.} of beams with small transverse extent. In round (cylindrically symmetric) geometry it is known that, for particles near the axis, the longitudinal wake is (nearly) independent of their transverse positions, and the dominant transverse wake---the dipole wake---depends linearly on the offset of the driving particle.  Consider, however, the flat geometry of two (longitudinally) corrugated infinite plates, with the aperture defined by $y=\pm a$. The transverse wakes and impedances, for particles near the axis, are of the form:~\cite{surface_impedance}
\begin{equation}
w_{y}(s)=y_{0}\,  w_{d}(s) + y\,  w_{q}(s)\ ,\quad w_x(s)= w_q(s)(x_0-x)\ ,
\end{equation}
with $(x_0, y_0)$ the transverse offset of the driving charge, $(x,y)$ that of the test charge.

 \subsection{Wakes at the Origin, at $s=0^+$}

For a periodic structure, the wake approaches a finite constant as $s\rightarrow0$, $w_0\equiv w(0^+)$, one that is independent of the properties of the boundary material. This has been shown to be true if the boundary is a resistive metal, a (metallic) disk-loaded structure, or a dielectric tube. It is probably generally true for boundaries made of metal or dielectric (see~\cite{Baturin}; but not if the boundary is a plasma~\cite{Gennady_plasma}). The constant $w_0$ depends only on the shape and size of the aperture ({\it e.g.} iris radius in disk-loaded RF structure) and on the (transverse) location of the particles. The same statement can be made about the slope of the transverse wakes at the origin $w_{y0}'$---for both the dipole and quad wakes discussed above.

For example, in a round structure, with the particles moving on axis, $w_0=4/ a^2$, with $a$ the radius of the aperture. However, for particles on axis in flat geometry, $w_0=\pi^2/(4a^2)$, where here $a$ represents the half-aperture of the structure.

One can see that, if one knows the impedance $Z(k)$, then $w_0$ is given by ($c/\pi$ times) the area under $Re[Z(k)]$. However, one can obtain $w_0$ also from the asymptotic behavior of $Z(k)$ at high frequencies~\cite{Gluckstern, zeroth_order}. The asymptotic impedance $Z_a(k)$ is related to the wake at the origin $w_0$. By letting the wake $w(s)\approx H(s)w_0$, we find the asymptotic form
    \begin{equation}
    Z_{a}(k)
    =
    \frac{w_0}{c}
    \int_{0}^\infty
    ds\, e^{iks}
    =
   i \frac{w_0}{kc}\ ,
    \end{equation}
    where we have neglected the contribution to the integral at the upper limit.
  Thus if we know $Z(k)$, we can obtain $w_0$ by the relation~\cite{zeroth_order}
  \begin{equation}
w_0=-ikcZ_{a}(k)=-ic\lim_{k\to\infty} kZ(k)\ ;\label{w0_eq}
\end{equation}
this procedure will yield a positive constant.

In the transverse case the wakes start at the origin linearly:
\begin{equation}
 w_y(s)= H(s)w'_{y0} s\ , 
\end{equation}
with $w'_{y0}$ the value of the slope of the wake at the origin. Substituting into the impedance formula, we find the form of the asymptotic impedance
    \begin{align}\label{eq:9}
    Z_{ya}(k)
    =
    -
    i\frac{w'_{y0}}{c}
    \int_{0}^\infty
    ds\,se^{iks}
    =
   i \frac{w'_{y0}}{ck^2}
    \ ,
    \end{align}
(where again we have neglected the contribution to the integral at the upper limit). Thus, from the high frequency impedance we obtain~\cite{zeroth_order}
\begin{equation}
w_{y0}'=-ik^2cZ_{ya}(k)=-ic\lim_{k\to\infty} k^2Z_{y}(k)\ ,\label{w0p_eq}
\end{equation}
which is a positive constant.

\section{The Concept of a Dechirper}

\begin{figure}[htb]
\centering
    \includegraphics[draft=false, width=.4\textwidth]{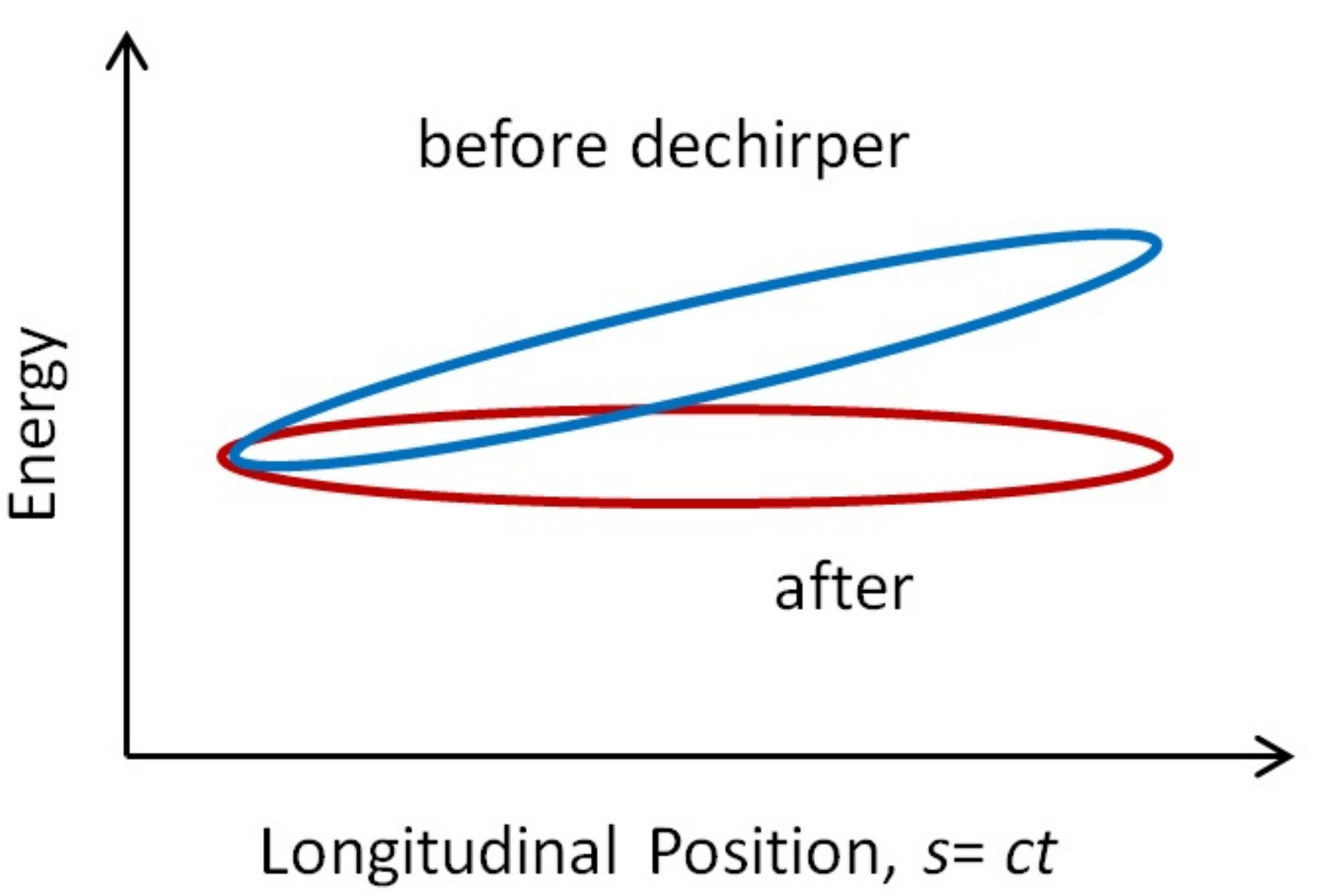}
\caption{ Sketch of typical longitudinal phase space of beam at the end of acceleration in a linac-based FEL, before the dechirper (blue), and after passing through the dechirper (red). The front of the bunch is to the left. }\label{phase_sketch_fi}
 \end{figure}

 In a linac-based, X-ray FEL, by the use of accelerating structures and chicanes, a low energy, low (peak) current beam ($\sim10$~MeV, $\sim100$~A) is converted to one with high energy and high current ($\sim5$--10~GeV, $\sim1$~kA).  After the last bunch compressor the beam is typically left with an energy--longitudinal position correlation (an energy ``chirp"), with the bunch tail at higher energy than the head (see Fig.~\ref{phase_sketch_fi}, the blue curve).

 A typical value of chirp might be $\nu=40$~MeV/mm. To cancel the chirp, one can run the beam off crest in downstream RF cavities. Running the beam on the zero crossing of the wave, we would need length $L_{rf}=\nu/(G_{rf}k_{rf})$ of extra RF. Or with peak RF gradient $G_{rf}=20$~MeV/m, wave number $k_{rf}=27$/m (for frequency $f=1.3$~GHz), we would need $L_{rf}=74$~m of extra active RF.
A dechirper is a passive way to achieve the same result in a few meters of structure.


 An ideal dechirper would have the wake: $w(s)=w_0H(s)$, with $H(s)=0$ (1) for $s<0$ ($s>0$).
 This is because at the end of an X-ray FEL the bunch distribution is (approximately) uniform: $\lambda(s)=H(s+\ell/2)H(\ell/2-s)/\ell$, with $\ell$ the full bunch length.
For an ideal dechirper and a uniform bunch distribution we find that Eq.~\ref{bunch_wake_eq} yields, $W_\lambda(s)=-w_0s/\ell$, which is linear in $s$, with the tail losing the most energy. The induced chirp, $\nu=-QLw_0/\ell$, with $Q$ bunch charge and $L$ the length of structure.
Note that a resistive beam pipe or periodic (passive) RF cavities do not work well as dechirpers, since the wake $w(s)$ drops quickly as $s$ moves away from the origin.

\subsection{Corrugated Structure as Dechirper}

\begin{figure}[htb]
\centering
\includegraphics[draft=false, width=.30\textwidth]{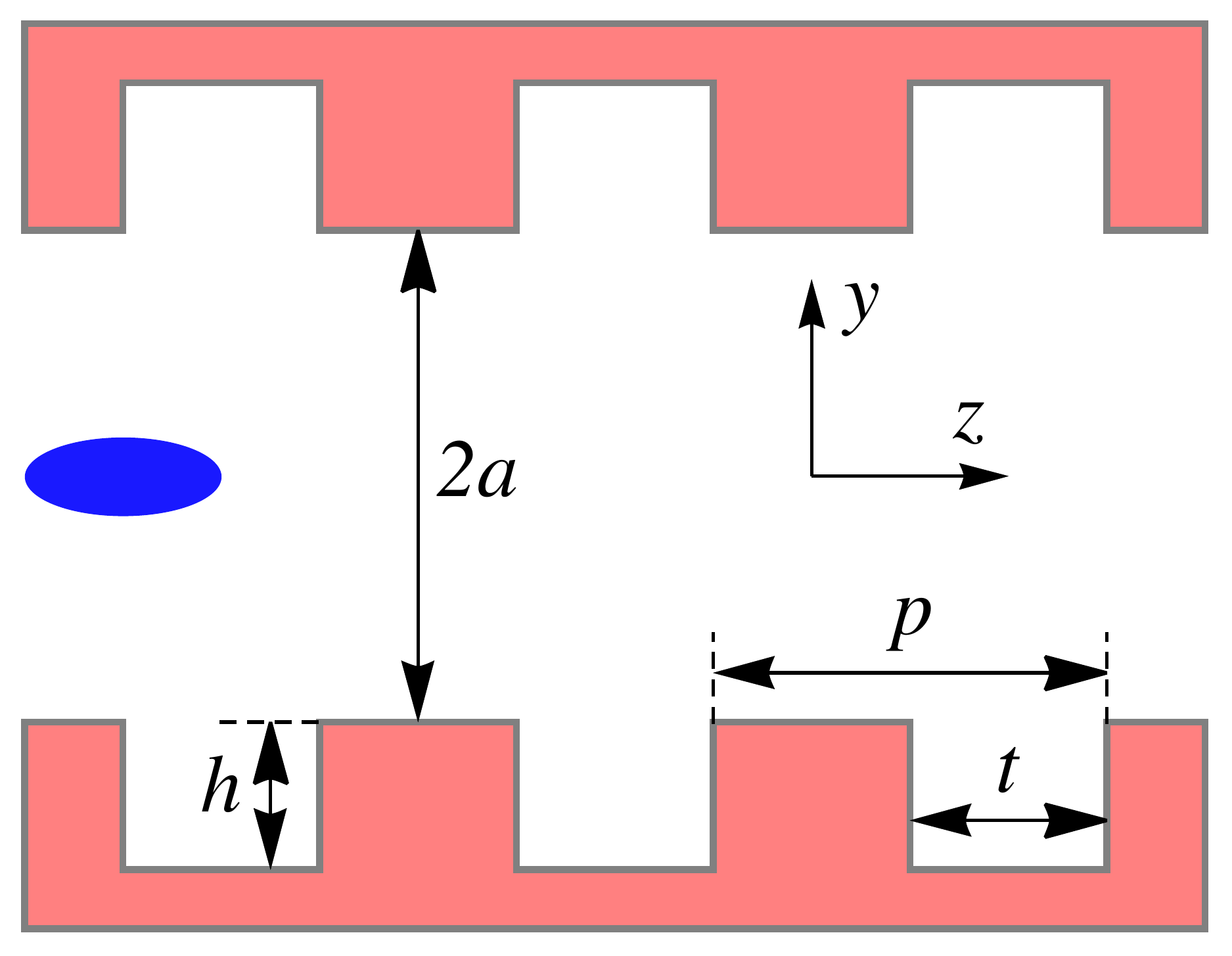}
\caption{Geometry of a dechirper showing three corrugations. The blue ellipse represents an electron beam propagating along the $z$ axis.  For the RadiaBeam/SLAC dechirper, (typical) half-gap $a=0.7$~mm, $h=0.5$~mm, $p=0.5$~mm, and $t=0.25$~mm.} \label{geometry_fi}
\end{figure}

A metallic pipe with small corrugations (see Fig.~\ref{geometry_fi}) can function as a good dechirper for an X-ray FEL~\cite{dechirper}. Ideally we would like $h,p\ll a$, and it is important that $h\gtrsim p$. If $h,p\ll a$, then perturbation solutions exist; in the case of round geometry,
the wake is dominated by a single mode: $w(s)\approx H(s)w_0\cos ks$, with $w_0=4/a^2$ and $k=\sqrt{2p/aht}$~\cite{round}.

\begin{figure}[!htb]
    \centering
    \includegraphics[draft=false, width=.45\textwidth]{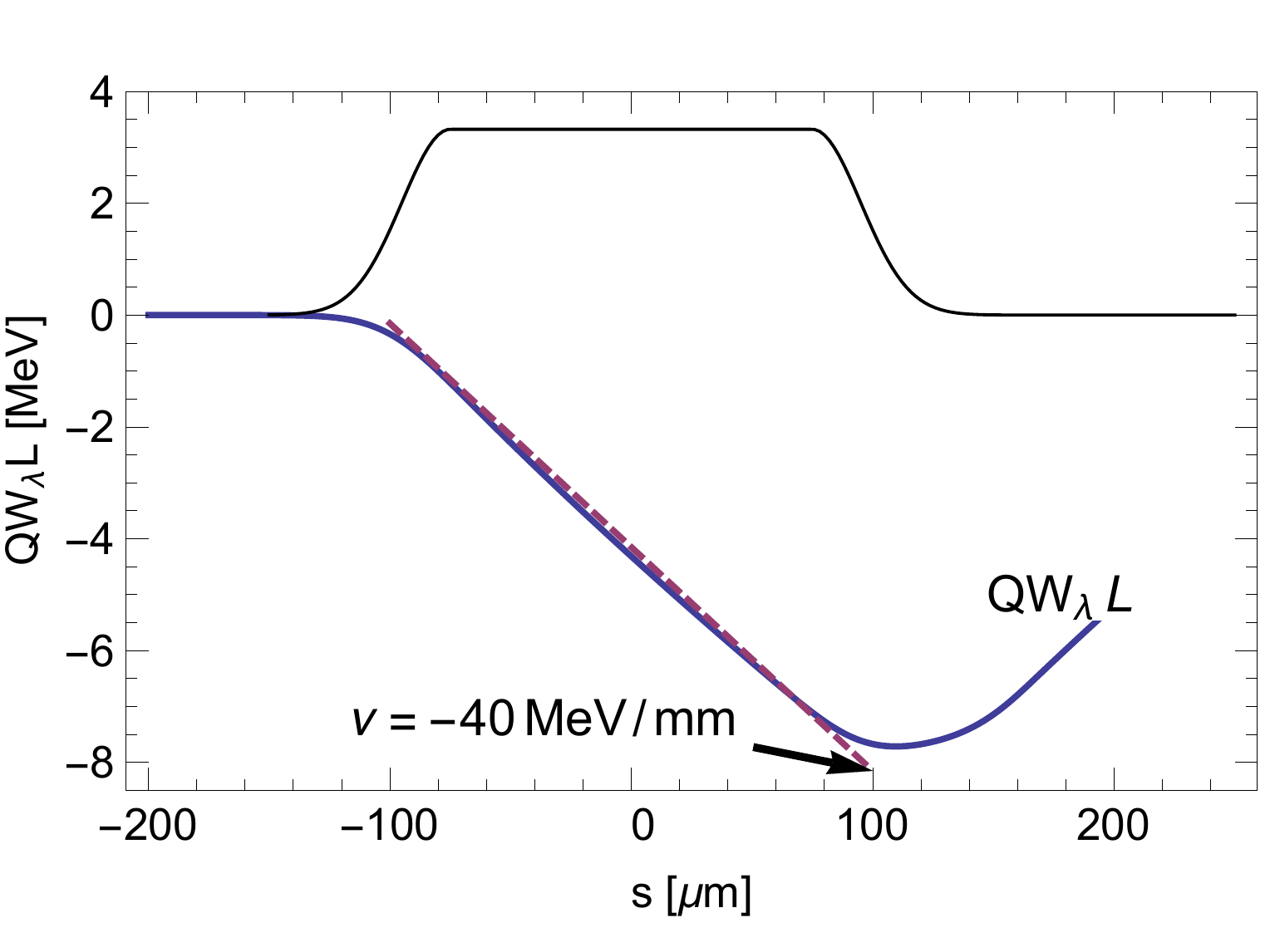}
\caption{ Dechirper for NGLS~\cite{dechirper}: wake of model of NGLS bunch distribution (blue). The dashed, red line gives the linear chirp approximation.
The bunch shape $\lambda$, with the head to the left, is given in black.}
\label{NGLS_fi}
    \end{figure}

As an example calculation we consider the beam properties of the NGLS project, where  $Q=300$~pC and $\ell=150$~$\mu$m.
We performed a time-domain wake calculation using I.~Zagorodnov's ECHO code~\cite{ECHO} for (round) dechirper parameters $a=3$~mm, $h=0.45$~mm, $p=1$~mm, $t=p/2$, and $L=8.2$~m. In Fig.~\ref{NGLS_fi} we show the numerical result (the blue curve), the analytical result (the dashed red line), and the bunch shape (in black, with the head to the left).  We see that the numerical result, over the core of the beam, agrees well with the analytical one.

Using a corrugated structure with flat geometry as dechirper has the advantage over round geometry in that the strength of interaction can be changed by simply changing the gap between the two plates. However, as we saw above, $w_0$ becomes weaker (for the same aperture). In addition, an unavoidable quad wake is excited, even when the beam is on axis; its effect, however, can simply be canceled by having half the dechirper oriented horizontally and half vertically. This, in fact, is the configuration of the RadiaBeam/SLAC dechirper that is installed in the LCLS.

\section{Surface Impedance Approach}

To find the impedance of a structure like the dechirper one can use the method of field matching (see {\it e.g.}~\cite{Zhen}). 
 However, if the corrugations are small  ($h\ll a$), the fields excited by the driving particle can be solved using a simpler method, a surface impedance approach~\cite{surface_impedance}. 
 According to this method, on the walls we let
\begin{equation}
\tilde E_z(k)=\zeta(k)\tilde H_x(k)\ ,
\end{equation}
with $\zeta(k)$ the surface impedance; and $E$, $H$, are the electric and magnetic fields. 
We solve Maxwell's equations (in the frequency domain) for the fields excited by the point driving charge. For flat geometry calculation we perform also the Fourier transform in $x$ of the fields, {\it e.g.} for the magnetic field we use
\begin{equation}
\hat H_x(q)=\int_{-\infty}^\infty dx\,\tilde H_xe^{iqx}\ .
\end{equation}
 We finally obtain the flat geometry {\it generalized impedances}, {\it i.e.} impedances where the transverse positions of driving and test particle can be located anywhere within the aperture.
 The longitudinal generalized impedance can be written in the form:~\cite{Bane}
\begin{equation}
 Z(x_0,y_0,x,y,k)=\int_{0}^\infty dq\,  f(q,y,y_0,k,\zeta)e^{-iq(x-x_0)}\ ,\label{Z_surf_eq}
\end{equation}
where $f$ is an explicit, analytical function of its arguments. The transverse generalized impedance is given in the same form, with a different (explicit analytical) function in the integral, $f_y(q,y,y_0,k,\zeta)$. Finally, the wake is obtained by numerically performing the inverse Fourier transform.
Thus, by performing two numerical integrals, we obtain an estimate of the generalized longitudinal and transverse wakes.
A subset of these results, one in which we are normally interested, is the special case of pencil beams, {\it i.e.} the case where $x\approx x_0$, $y\approx y_0$.


 For the case of a beam passing near a single plate, we begin with the impedance for two plates separated by distance $2a$. In the expression for $f(q,y,y_0,k,\zeta)$ described above we let $y=a-b$ and then $a\rightarrow\infty$. Using the new version of $f$ and performing the same integrals as before, we obtain the wakes of a beam passing by a single dechirper jaw at distance $b$.

 The only problem with using the surface impedance approach for the calculation of the RadiaBeam/SLAC dechirper is that it normally is valid only if $(h/a)\ll1$, whereas here nominally $(h/a)=(0.5/0.7)=0.7\not\ll1$ (and similarly for the single jaw case). 
 However, for the short, LCLS-type bunches---with rms length of 10's of microns---we demonstrate below that this approach still works when used with a surface impedance that represents the wall corrugations at high frequencies.
 


\section{Explicit Analytical Approximations of Wakes}

We have further simplified the results by extracting (analytical) parameters and using simplified formulas for cases of pencil beams. In the {\it zeroth order} approximation~\cite{zeroth_order}, we let $w(s)=H(s)w_0$, where $w_0$ is obtained using Eqs.~\ref{w0_eq}, \ref{Z_surf_eq} (longitudinal case), and $w_y(s)=H(s)w_{y0}'s$, where $w_{y0}'$ is obtained using Eqs.~\ref{w0p_eq}, \ref{Z_surf_eq} (using $f_y$; transverse case).

 For better agreement with results obtained directly from Eq.~\ref{Z_surf_eq}  and with numerical simulations, we use the {\it first order} approximation~\cite{Bane}, which in the longitudinal case is of form
\begin{equation}
w(s)=H(s)w_0e^{-\sqrt{s/s_0}}\ .\label{wa_eq}
\end{equation}
 Parameter $s_0$ can also be derived in analytical form from the structure of the impedance at high frequencies. Eq.~\ref{wa_eq} corresponds to the two-term, high-frequency Taylor expansion:
 \begin{equation}
Z(k)\approx i\frac{w_0}{kc}\left[1-\frac{(1+i)}{\sqrt{2ks_{0}}}\right]\quad\quad (k\rightarrow\infty)\ .\label{imp_round_eqb}
\end{equation}
Thus, to obtain parameter $s_0$, we begin with general form of the impedance (Eq.~\ref{Z_surf_eq}), and substitute in the high frequency surface impedance for the corrugated structure:~\cite{Gennady_periodic,Bane} 
\begin{equation}
\zeta(k)=\frac{1}{\alpha p}\left(\frac{2it}{\pi k}\right)^{1/2}\ ,
\end{equation} 
with $\alpha\approx1-0.465\sqrt{t/p}-0.070(t/p)$. 
We then expand to two terms in Taylor series (at high $k$), integrate over $dq$, and find $s_0$ by comparing with the form of Eq.~\ref{imp_round_eqb}.
A similar procedure is followed for the transverse (dipole and quad) wakes; in this case, the wake is of the form
\begin{equation}
w_{y}(s)=
     2H(s)w_{0y}'
   s_{0y}\left[1-\left(1+\sqrt{\frac{s}{s_{0y}}}\right)e^{-\sqrt{s/s_{0y}}}\right]\ ,\label{wya_eq}
\end{equation}
with parameters $w_{0y}'$ and $s_{0y}$. 
Note that the analytical forms for the longitudinal and transverse, short-range wakes (Eqs.~\ref{wa_eq}, \ref{wya_eq}) have been used before, for the case of periodic, disk-loaded accelerating structures~\cite{nlc_wake, nlc_wakex}; a major difference, however, was that the parameters $s_0$, $s_{0y}$, were there obtained through fitting to numerical results.

\begin{figure}[!htb]
    \centering
    \includegraphics[draft=false, width=.45\textwidth]{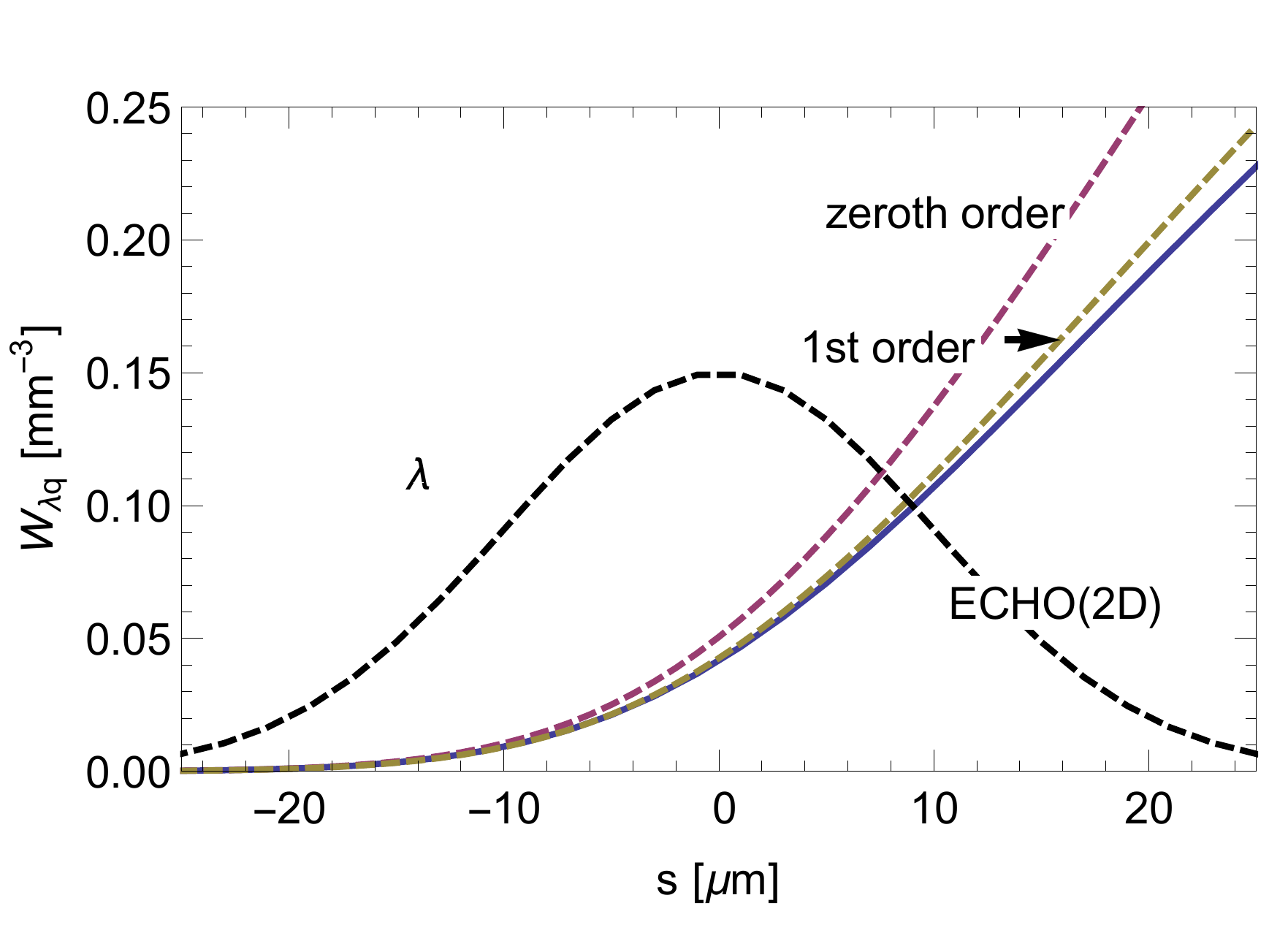}
\caption{ Quad bunch wake for a Gaussian beam on axis, with $\sigma_z=10$~$\mu$m, $a=0.7$~mm, in the RadiaBeam/SLAC dechirper~\cite{Bane}. Given are the numerical results of ECHO(2D) (blue), and the analytical zeroth order (red) and 1st order results (green). The bunch shape $\lambda(s)$ is shown in black.}\label{two_plate_ex_fi}
    \end{figure}

We now give results for two specific example calculations. In the first example we consider a short beam on the axis of a two-plate dechirper and find the quad wake, where the two analytical wake parameters are given by~\cite{Bane}
\begin{equation}
w_{0q}'= \frac{\pi^4}{32a^4}\ ,\quad s_{0q}=\frac{a^2t}{2\pi\alpha^2p^2}\left(\frac{15}{16}\right)^2\ .
\end{equation}
We consider the RadiaBeam/SLAC dechirper parameters with half-gap $a=0.7$~mm, and a Gaussian driving bunch with $\sigma_z=10$~$\mu$m. We obtain the zeroth and first order analytical bunch wakes by performing the convolution Eq.~\ref{bunch_wake_eq}. The analytical and numerical results of ECHO(2D)~\cite{echo2d} are shown in Fig.~\ref{two_plate_ex_fi}.  We see that the agreement of the 1st order analytical result and the numerical one is quite good.

\begin{figure}[!htb]
    \centering
    \includegraphics[draft=false, width=.45\textwidth]{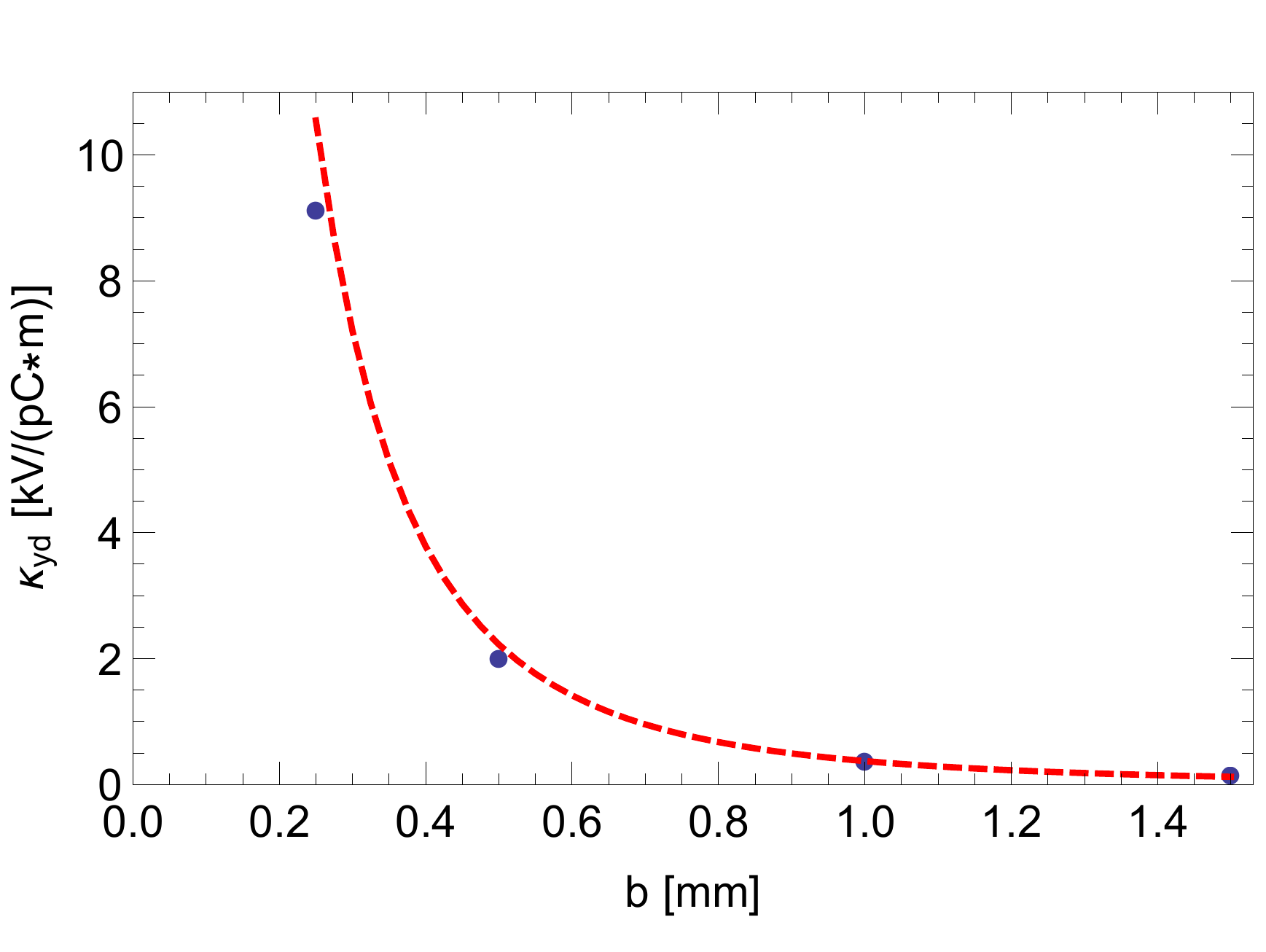}
\caption{ Single plate dipole kick factor $\varkappa_{yd}$ as function of distance of the beam from the wall $b$, showing the CST results (blue symbols) and those of the 1st order analytical model (red dashes)~\cite{single_plate_confirm}. The bunch is Gaussian with length $\sigma_z=100$~$\mu$m.}\label{one_plate_ex_fi}
    \end{figure}

Our second example considers the dipole wake of a beam offset by distance $b$ from one plate. For this case the wake parameters are given by~\cite{single_plate}
\begin{equation}
w_{0d}'= \frac{1}{b^3}\ ,\quad s_{0d}=\frac{8b^2t}{9\pi\alpha^2p^2}\ .
\end{equation}
In Fig.~\ref{one_plate_ex_fi} we plot, as functions of $b$, the analytically obtained (1st order) kick factor $\varkappa_{yd}$---the average of the bunch wake $W_{\lambda d}(s)$---and the numerical result obtained using CST Studio~\cite{CST}. The bunch here is Gaussian with length $\sigma_z=100$~$\mu$m. We see from the figure that the agreement of the analytical and numerical calculations is very good.

\vspace{-8mm}\section{Measurements}

\begin{figure}[!htb]
    \centering
    \includegraphics[draft=false, width=.45\textwidth]{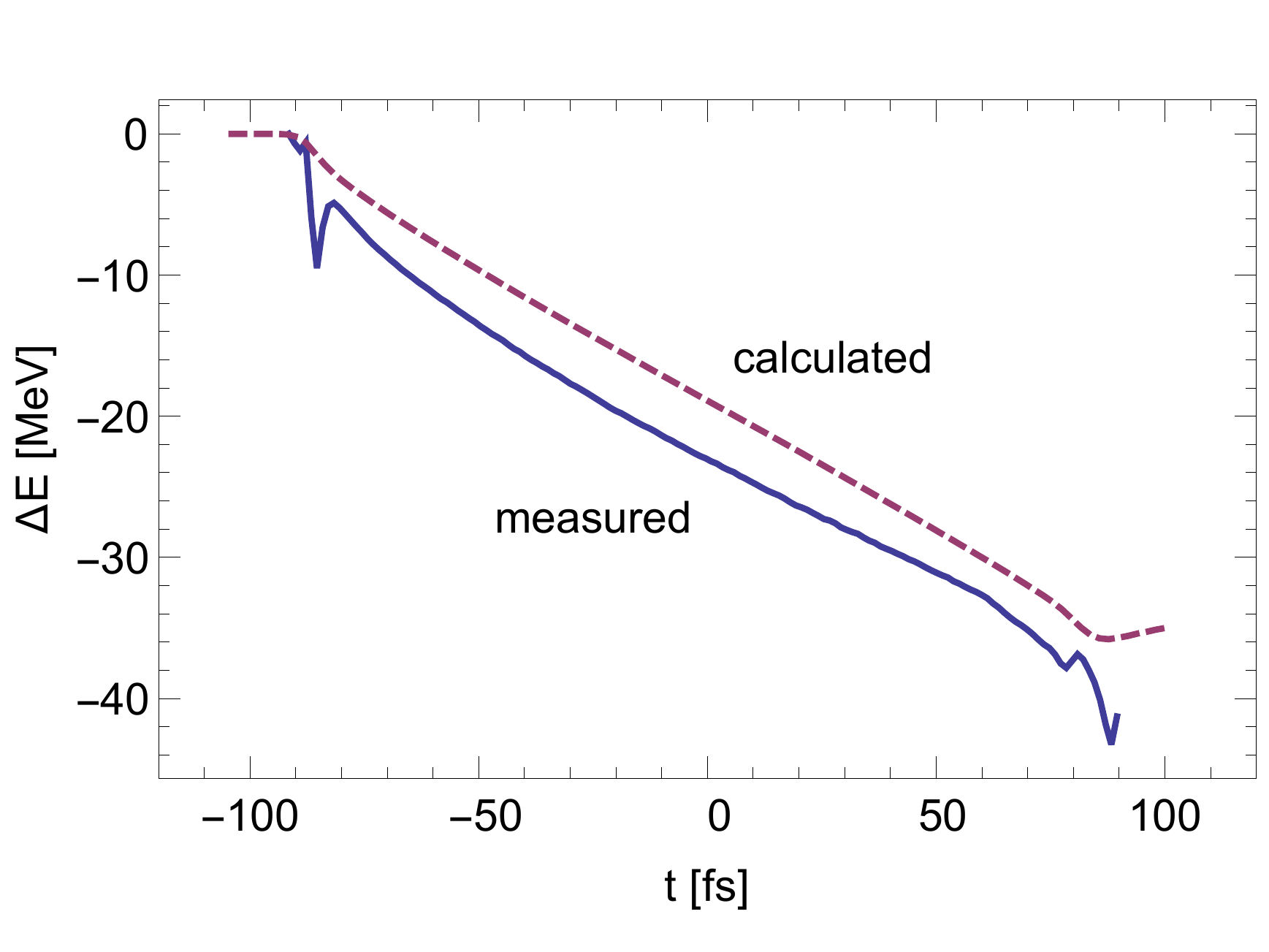}
\caption{Chirp induced in beam passing on axis between the jaws of both dechirper modules, according to measurement (blue) and calculation (red dashes)~\cite{MarcG}. Here $Q=190$~pC, $I=1.0$~kA, $E=4.4$~GeV. The bunch head is to the left.}
\label{dechirp_fi}\vspace{-6mm}
    \end{figure}

One example measurement of the longitudinal effect of the beam on axis between the jaws of both dechirper modules is shown in Fig.~\ref{dechirp_fi}. Here $Q=190$~pC, energy $E=4.4$~GeV, and dechirper half gap $a=1.2$~mm. The X-band, deflecting cavity diagnostic, XTCAV, measures longitudinal phase space of the bunch after the undulator in the LCLS. From it, we obtain the longitudinal bunch distribution $\lambda(s)$ and the induced energy chirp $\Delta E(s)$ due to the dechirper, by taking the average energy at position $s$ minus that when the dechirper jaws are wide open. Here the bunch shape is approximately uniform, with peak current $I=cQ\lambda=1.0$~kA.  For the calculation we convolve the measured $\lambda(s)$ with the analytical $w(s)$ (see Eq~\ref{bunch_wake_eq}), noting that $w_0=\pi^2/(4a^2)$ and $s_0=9a^2t/(8\pi\alpha^2p^2)$; then $\Delta E=eQW_\lambda L$, with $L=4$~m, the length of dechirper. 
In Fig.~\ref{dechirp_fi} the measured chirp using XTCAV (with arbitrary horizontal and vertical offsets) is given in blue, the calculation (with the head of the bunch at $t\equiv s/c=-85$~fs) is given in red dashes. We see that the agreement in induced chirp is good.

\begin{figure}[!htb]
    \centering
    \includegraphics[draft=false, width=.45\textwidth]{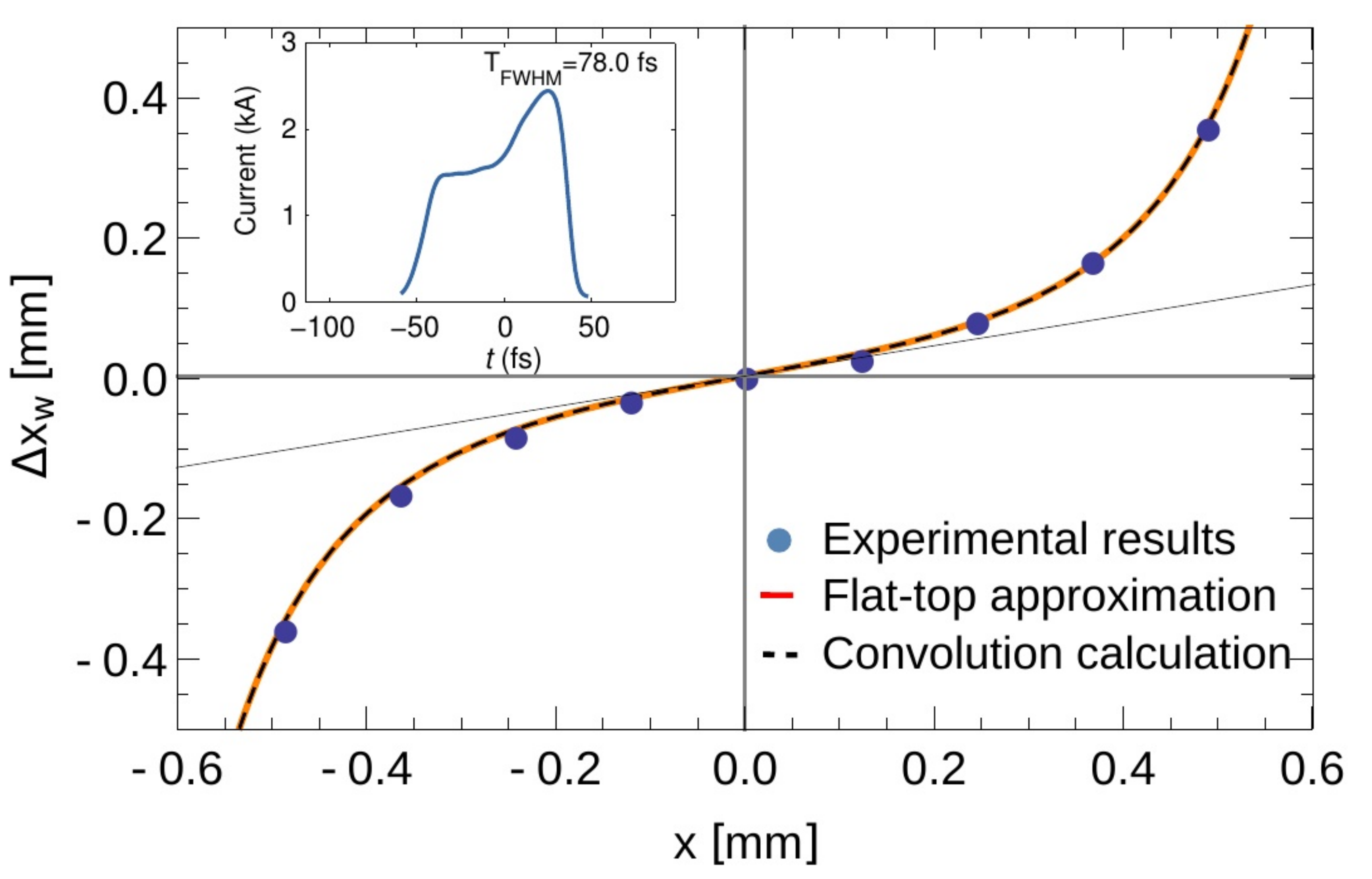}
\caption{Downstream deflection, $\Delta x_w$, as function of offset in the (horizontal) dechirper, $x$, for half-gap $a=1$~mm~\cite{Zemella}. Here $Q=152$~pC, $E=6.6$~GeV. For the analytic curves, the gap parameter was reduced by 11\% to fit the experimental data.
The bunch current, with head to the left, is shown in the inset.}\label{two_plates_a1mm_fi}\vspace{-2mm}
    \end{figure}

In the next example the beam was kept fixed, half the dechirper (the horizontal half) was scanned across the beam (keeping the half-gap $a$ fixed), while the other half (the vertical half) was opened wide, and the transverse kick was measured. The parameters are half gap $a=1$~mm, bunch charge $Q=152$~pC, and energy $E=6.6$~GeV.  The results are given in Fig.~\ref{two_plates_a1mm_fi}, where we plot the  offset at a downstream BPM, $\Delta x_w$ {\it vs.} offset of the beam from the axis in the dechirper, $x$. The figure shows the data (plotting symbols) and the calculations (the curves). 
The bunch shape as obtained by measurement (with head to the left), is given in the inset plot. 
For the scale of the wake effect, note that a value of $\Delta x_w=0.4$~mm corresponds to an average kick of $V_x=160$~kV.

 For the comparison calculations, we first numerically performed the convolution of Eq.~\ref{bunch_wake_eq} to obtain the bunch wake; then the result is given by $\Delta x_w=eQW_{\lambda x}L_{BPM}/E$, with $L_{BPM}$ ($=16.26$~m) the distance between dechirper and measuring BPM (the dashed curve in the figure). For the analytic curves, the gap parameter was reduced by 11\% to fit the experimental data. 
 The agreement between theory and measurements is good; the discrepancy in scale is small. In aligning the structure, the ends of the plates are independently adjusted; one possible cause of the discrepancy is that, during measurement, the jaws have an unknown residual tilt. 
 


\begin{figure}[!htb]
    \vspace{-4mm}
    \centering
    \includegraphics[draft=false, width=.45\textwidth]{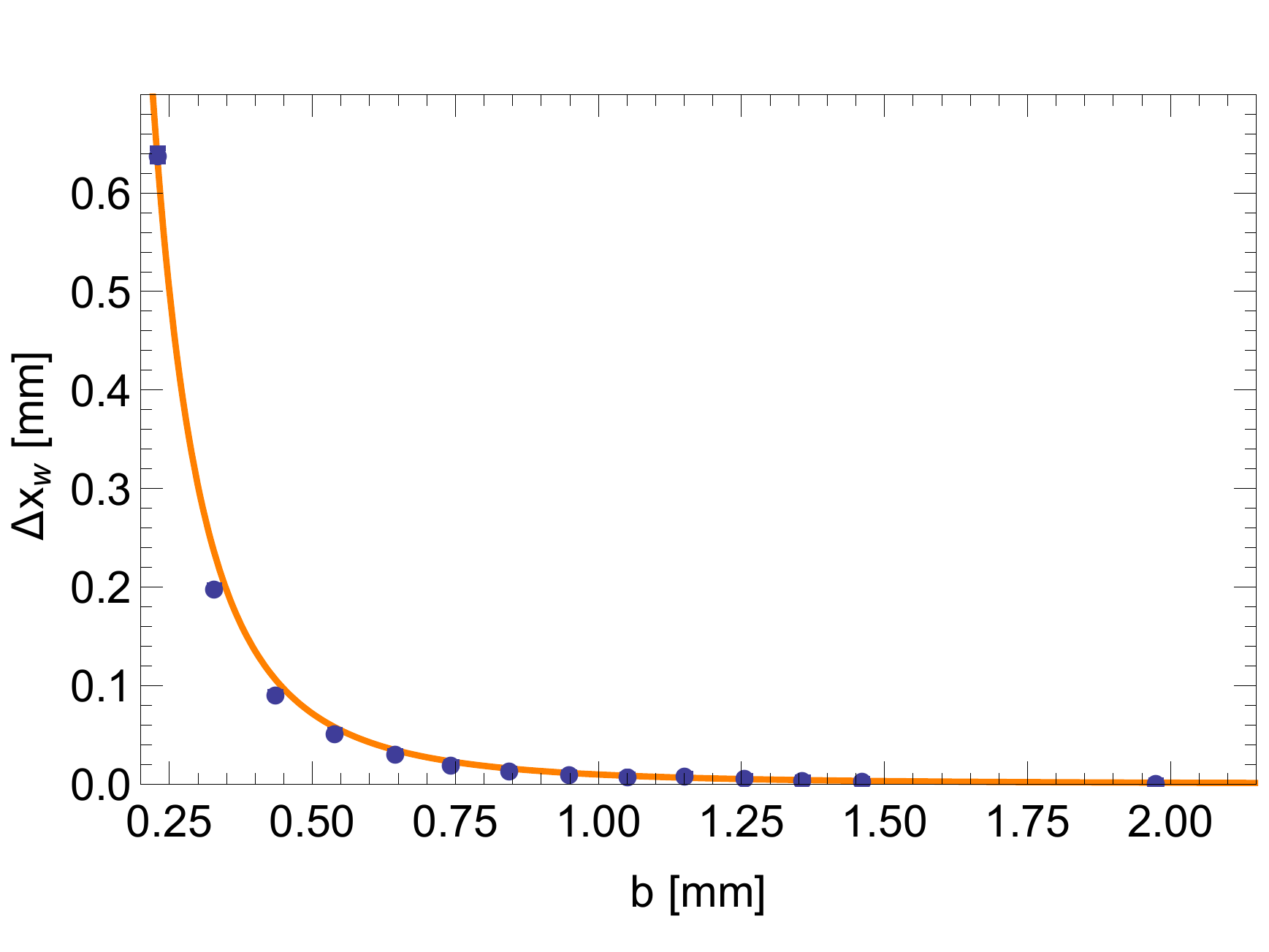}
\caption{Downstream deflection, $\Delta x_w$ as functions of beam offset from a single dechirper plate, $b$~\cite{Zemella}. Here $Q=180$~pC, $I=3.5$~kA, $E=13$~GeV.  The symbols give the data points, with their $b$ values shifted by $-161$~$\mu$m; the curves give the analytical theory. }\label{north_kick_fi}
\vspace{-2mm}
    \end{figure}

The final example is a single jaw measurement of downstream offset due to transverse kick,  $\Delta x_w$ {\it vs.} beam offset in dechirper, $b$ (see  Fig.~\ref{north_kick_fi}). Note that $\Delta x_w>0$ indicates a kick toward the jaw, and a value of $\Delta x_m=0.6$~mm corresponds to a kick of $V_x=480$~kV.
 The beam parameters were charge $Q=180$~pC, peak current $I=3.5$~kA,
  and energy $E=13$~GeV. 
  The absolute offset of the beam from the dechirper plate is not well known.
Therefore, a fit of the model to the measurement was performed, where an overall shift in beam offset $\Delta b$ was allowed.
We see that the fit of the theory (the curve) to the data (the plotting symbols) is good. The fit gives an overall shift of $\Delta b=-161$~$\mu$m, and the data in the plot has been shifted by this amount.
During this measurement, the longitudinal wake effect was simultaneously recorded, using a downstream BPM at a location of dispersion. The agreement between the longitudinal theory and measurement was again good, when an overall shift of $\Delta b=-138$~$\mu$m was assumed in the theory. The fact that, in both cases, the theory and measurement curves agree well, and that the fitted shifts are close to each other, is confirmation of our wake models and of our analysis.

\vspace{-7mm}\section{Summary}

 The corrugated, metallic structure can be used for passive chirp control at the end of a linac-based, X-ray FEL.
It can also be used as a fast kicker, to facilitate two-color, fresh-slice operation of an FEL such as the Linac Coherent Light Source (where it is regularly being used for this purpose).
Using the surface impedance approach, we are able to obtain analytical solutions for the wakes of the structure: longitudinal, dipole, quad wakes; two-plate case, on axis and off; and single plate case.
These wakes agree well with numerical simulations for LCLS-type beam parameters, in spite of the fact that the corrugation perturbation is often not small. They also agree quite well with measurements performed using the RadiaBeam/SLAC dechirper at the LCLS.

More comparisons with numerical simulations can be found in \cite{Bane, single_plate_confirm}, with measurements in \cite{MarcG, Zemella}.

\end{document}